\documentclass[12pt]{article}

\usepackage[english]{babel}
\usepackage[latin1]{inputenc}
\usepackage{times}
\usepackage[T1]{fontenc}

\usepackage{etex}
\usepackage{amsmath,amssymb,amscd,mathtools,array}
\usepackage{graphics,graphicx,xcolor,txfonts}
\usepackage[all]{xy}
\usepackage{fancybox}

\usepackage{color}
\usepackage[colorlinks=true, pdfstartview=FitV, linkcolor=blue, citecolor=blue, urlcolor=blue]{hyperref}

\let\ssection=\section
\renewcommand{\section}{\setcounter{equation}{0}\ssection}

\newcommand{\bA}{{\bf A}}
\newcommand{\tbA}{\widetilde{\bf A}}

\newcommand{\bb}{{\bf b}}
\newcommand{\bB}{{\bf B}}

\newcommand{\bbeta}{\boldsymbol{\beta}}
\newcommand{\bone}{\boldsymbol{1}}

\newcommand{\bC}{{\bf C}}
\newcommand{\bbC}{\mathbb{C}}

\newcommand{\Div}{\mathrm{Div}}

\newcommand{\be}{\mathbf{e}}
\newcommand{\barbe}{\overline{\be}}

\newcommand{\barE}{\overline{E}}
\newcommand{\bE}{{\bf E}}

\newcommand{\cF}{{\mathcal{F}}}

\newcommand{\bg}{{\bf g}}

\newcommand{\rg}{\mathrm{g}}

\newcommand{\fg}{\mathfrak{g}}

\newcommand{\grad}{{\mathbf{grad}}}
\newcommand{\bargamma}{\overline{\gamma}}

\newcommand{\fh}{\mathfrak{h}}

\newcommand{\Id}{\mathrm{Id}}

\newcommand{\sfJ}{\mathsf{J}}

\newcommand{\bell}{\boldsymbol{\ell}}

\newcommand{\cM}{\mathcal{M}}

\newcommand{\tcM}{\widetilde{\cM}}

\newcommand{\bn}{{\bf n}}

\newcommand{\bbN}{\mathbb{N}}

\newcommand{\cO}{\mathcal{O}}

\newcommand{\bomega}{\boldsymbol{\omega}}

\newcommand{\tomega}{\widetilde{\omega}}

\newcommand{\bp}{{\bf p}}

\newcommand{\bP}{{\bf P}}
\newcommand{\cP}{\mathcal{P}}
\newcommand{\barpsi}{\overline{\psi}}
\newcommand{\bpi}{\boldsymbol{\pi}}

\newcommand{\tvarpi}{\widetilde{\varpi}}

\newcommand{\bq}{{\bf q}}

\newcommand{\bbR}{\mathbb{R}}

\newcommand{\rot}{\mathbf{rot}}

\newcommand{\bs}{{\bf s}}

\newcommand{\cS}{{\mathcal{S}}}
\newcommand{\sign}{\mathrm{sign}}
\newcommand{\SE}{\mathrm{SE}}

\newcommand{\SO}{\mathrm{SO}}
\newcommand{\se}{\mathfrak{se}}
\newcommand{\so}{\mathfrak{so}}
\newcommand{\su}{\mathrm{su}}
\newcommand{\SU}{\mathrm{SU}}

\newcommand{\bsurf}{\mathbf{surf}}

\newcommand{\bsigma}{\boldsymbol{\sigma}}

\newcommand{\tsigma}{\widetilde{\sigma}}
\newcommand{\sfS}{\mathsf{S}}

\newcommand{\ts}{{\tilde{s}}}

\newcommand{\Tr}{\mathrm{Tr}}

\newcommand{\bu}{{\bf u}}
\newcommand{\barbu}{\bar{\bu}}

\newcommand{\rU}{\mathrm{U}}

\newcommand{\bv}{{\bf v}}

\newcommand{\cV}{{\mathcal{V}}}
\newcommand{\hcV}{{\widehat{\cV}}}
\newcommand{\tcV}{{\widetilde{\cV}}}

\newcommand{\Vol}{\mathrm{Vol}}

\newcommand{\bw}{{\bf w}}

\newcommand{\bx}{{\bf x}}

\newcommand{\barbx}{\bar{\bx}}

\newcommand{\bz}{\mathbf{z}}

\newcommand{\barbz}{\overline{\bz}}

\newcommand{\bbZ}{\mathbb{Z}}

\newcommand{\bzeta}{\boldsymbol{\zeta}}
\newcommand{\barbzeta}{\overline{\bzeta}}

\newcommand{\const}{\mathop{\rm const.}\nolimits}
\newcommand{\half}{\frac{1}{2}}

\newcommand{\la}{{\langle}}
\newcommand{\ra}{{\rangle}}

\newcommand{\bigbox}[1]{\fbox{%
\rule[-20pt]{0pt}{45pt}$\;\;\displaystyle{#1}\;\;$}%
}
\newcommand{\medbox}[1]{\fbox{%
\rule[-15pt]{0pt}{35pt}$\;\;\displaystyle{#1}\;\;$}%
}

\newcommand{\blue}[1]{{\textcolor{blue}{#1}}}
\newcommand{\red}[1]{{\textcolor{red}{#1}}}

\begin{document}

\baselineskip=19pt
\oddsidemargin .1truein

\newtheorem{thm}{Theorem}[section]
\newtheorem{lem}[thm]{Lemma}
\newtheorem{cor}[thm]{Corollary}
\newtheorem{pro}[thm]{Proposition}
\newtheorem{ex}[thm]{Example}
\newtheorem{rmk}[thm]{Remark}
\newtheorem{defi}[thm]{Definition}

\newenvironment{dedication}
    {\vspace{6ex}\begin{quotation}\begin{center}\begin{em}}
    {\par\end{em}\end{center}\end{quotation}}

\title{{\sc{}Polarized Spinoptics\\ and \\ Symplectic Physics}}

\author{
Christian DUVAL\footnote{mailto: duval@cpt.univ-mrs.fr}\\[6pt]
Aix-Marseille Universit\'e, CNRS, CPT, UMR 7332, 13288 Marseille, France.\\
Universit\'e de Toulon, CNRS, CPT, UMR 7332, 83957 La Garde, France.
}

\date{\today}

\maketitle

\thispagestyle{empty}

\begin{dedication}
To 
Jean-Marie Souriau, with admiration and faithfulness
\end{dedication}

\begin{abstract}
We recall the groundwork of spinoptics based on the coadjoint orbits, of given color and spin, of the group of isometries of Euclidean three-space; this model has originally been put forward by Souriau in his treatise \textsl{Structure des Syst\`emes Dynamiques}, whose manuscript was initially entitled \textsl{Physique symplectique}.
We then set up a model of polar\-ized spinoptics, namely an extension of geometrical optics accounting for elliptical\-ly polar\-ized light rays in terms of a certain fibre bundle associated with the bundle of Euclide\-an frames of a given Riemannian three-manifold. The characteristic foliation of a natural pre\-symplectic two-form introduced on this bundle via the Ansatz of minimal coupling is determined, yielding a set of differential equations governing the trajectory of light, as well as the evolution of polarization in this Rieman\-nian manifold. Those equations, when specialized to the Fermat metric (for a slowly varying refractive index), enable us to recover, and justify, a set of differential equations earlier proposed in the literature, in another context, namely in terms of a semi-classical limit of wave optics. They feature a specific anomalous velocity res\-ponsible for the recent\-ly observed Spin Hall Effect of Light, namely a tiny spatial deflection of polarized light rays, transversal\-ly to the gradient of the refractive index. Our model, constructed from the start on purely geometric grounds, turns out to encode auto\-matically the Berry as well as the Pancharatnam connections that usually appear in the framework of wave optics. 
\end{abstract}

\newpage
\tableofcontents

\section{Introduction}

Geometrical optics, going back in time as early as Euclid's Optika \cite{Euc},
has a very long history pinnacling with the discoveries of the laws of reflection and refraction by Ibn Sahl (984), Snel (1621), and Descartes \cite{Des}. But the main step 
has been taken by Fermat~\cite{Fer} via the ``principle of least optical path'', opening the way to the modern calculus of variations. 
Geometrical optics has, since then, been at the heart of the focal interplay between geo\-metry and physics, see, e.g., \cite{Car,Sou,Arn,AI,BMD,BNKH,Nor}; its past and present is multi-faceted and rich enough to deter anyone from undertaking a detailed survey. We will, henceforth, limit ourselves to review some of the most recent developments of what has been coined \textit{spinoptics} and its generalizations.

From the mid-sixties, onward, Souriau has highlighted the distinguished status of homo\-geneous symplectic manifolds of a given Lie group, $G$, of ``space(time) symmetries'' that constitute the classical \textit{elementary systems}, some of them  corresponding to unitary ir\-reducible representations of $G$, interpreted via Geometric Quantization~(GQ) as the quantum counterparts of these classical physical systems~\cite{Sou}. This correspondence has also been independently put forward, in the context of representation theory, by Kirillov~\cite{Kir} and Kostant~\cite{Kos}. As regards symplectic physics, Souriau has achieved the classifica\-tion of (prequantizable) homogeneous symplectic manifolds for the Galilei and Poincar\'e groups; he has shown that GQ indeed leads to the corresponding (free) quantum wave equations, e.g., to the Schr\"odinger and Schr\"odinger-Pauli equations in the non-relativistic case, and to the Klein-Gordon, Dirac, Maxwell equations in the relativistic framework.

Now, the coadjoint orbits of the Euclidean group, $\SE(3)$, have been completely classified by Souriau \cite{Sou}, see also \cite{GS,MR}. Those with spin Casimir $s=\pm\hbar$, i.e., the elementary classical states of spinning light rays, have played a paramount r\^ole in the theory of \textit{spinoptics}. These Euclidean coadjoint orbits have recently been taken in due consideration \cite{DHH1,DHH2,Duv} in an effort to provide a purely classical interpretation of the subtle Spin Hall Effect of Light (SHEL) \cite{BB04,OMN}. Let us emphasize that the SHEL, a \textit{tiny} spin-governed transverse shift of light rays across an optical interface, has been observed experimental\-ly~\cite{HK,BNKH} very recent\-ly using the innovative method of ``weak quantum measure\-ments'' --- we refer to the latest, comprehensive, review of ``Beam Shifts'' written by Bliokh and Aiello \cite{BA}. We have indeed shown \cite{DHH1,DHH2,DHZ} that, astonishingly enough, the SHEL admits a classical interpretation in terms of  Souriau's symplectic scattering \cite{Sou} between $\SE(3)$-coadjoint orbits modeling classical photonic states. Let us mention, \textit{en passant}, that spinoptics has also been formulated on Finsler manifolds~\cite{Duv} to cope with inhomogeneous and anisotropic optical media.

\goodbreak

The purpose of this article is to enlarge the previous approach, which dealt essentially with circular polarization, so as to incorporate, in an exclusively geometric fashion, the generic elliptic polarization of light. Our endeavor is therefore to set up a $\SE(3)$-invariant (pre)symplectic model so as to make it possible to account for the geo\-metrodynamics of polar\-ized light rays in inhomogeneous dielectric media; our standpoint is therefore to build our theory starting with the Euclidean photonic models rather than from a semi-classical approximation of stationary Maxwell's field equations, as usual in the optics literature \cite{LZ}.

The plan is as follows.

We recall in Section \ref{Spinoptics} Souriau's definition of an elementary system associated with a coadjoint orbit of a given space(time) symmetry group $G$. This definition, once specialized to the Euclidean group, $G=\SE(3)$, leads to the theory spinoptics (in vacuum) where the manifold of spinning light rays is viewed as the coadjoint orbit of a given point $\mu_0\in\se(3)^*$ with color, $p$, and spin,~$s$. It is well-known that the prequantizable orbits are associated with~$s\in\hbar\bbZ$; those for which $s=\pm\hbar$ characterize \textit{photons} as advocated by their geometric quantization which, as we show, lead to the stationary Maxwell equations for \textit{Euclidean wave-optics}. Spinoptics merely describes classical states of circularly polarized photons. 

In an attempt to go a step further and aim at a classical description of polarized light rays, we resort, in Section \ref{PolarizedLightSection}, to a natural Hermitian extension $\widetilde{\strut\SE(3)}=\SU(3)\ltimes\bbC^3$ of the Euclidean group to work out Souriau's construction of a Hamiltonian $\SE(3)$-space, namely the manifold~(\ref{tcM}) endowed with the symplectic $2$-form (\ref{omegaBis}) associated with the previous photonic origin $\mu_0$. See Figure \ref{Diagram} for a diagrammatic rendition. The new observable superseding the photonic spin for polarized light is then the $\rU(1)$-momentum mapping,~$\ts$, given by (\ref{s3Bis}); see also~(\ref{J=sBis}). Let us emphasize that the natural $1$-form (\ref{tvarpi}) accordingly defined on the \textit{evolution space} (\ref{tVbis}) features in an unifying way the Fermat $1$-form, the Berry and the Pantcharatnam connection forms, as conspicuous from the expression (\ref{varpiTer}).

Section \ref{InHomogeneousSection} is ultimately devoted to the determination of the system of ordinary dif\-ferential equations governing the dynamics of polarized light rays in inhomogeneous iso\-tropic dielectric media modeled by Riemannian $3$-dimensional connected manifolds $(M,\rg)$. This system stems, in full generality, from the characteristic foliation of a new \textit{evolution space}~$(\tcV,\tsigma)$ defined by (\ref{cV}), and (\ref{varpiNew}), via the prescription of ``minimal coupling'' to the Levi-Civita connection of a given metric, $\rg$, on configuration space $M$. The resulting differential equations happen to generalize those obtained by Liberman and Zel'dovich~\cite{LZ}. We  duly recover the latter equations in the regime of a slowly varying refractive index defining a Fermat metric on Euclidean space $E^3$. In particular, we do confirm the --- quite recently experimentally verified \cite{BNKH} --- precession of the Stokes vector which accounts for the polarization state of the system in such an optical medium. This section ends with the derivation of the above-mentioned \textit{Spin Hall Effect of Light} in the herewith developed formalism of \textit{polarized spinoptics}.

\subsubsection*{Acknowledgements}
A great many thanks are due to P.~Horv\'athy, P.~Iglesias-Zemmour, and F.~Ziegler for enlight\-ening and friendly discussions during the preparation of this article dedicated to the memory of our teacher Jean-Marie-Souriau. It is also a pleasure to acknowledge the interest that K.~Bliokh has always kindly showed for our geometrical approach to optics. 

\goodbreak

\section{An overview of Euclidean spinoptics}\label{Spinoptics}

We recall, and justify, that the $\SE(3)$-coadjoint orbits of color $p$ and spin $s=\pm\hbar$ may be interpreted as providing a model of \textit{circularly polarized} Euclidean photon \cite{Sou}, that serves as a fundamental ingredient in the theory of spinoptics \cite{DHH1,DHH2}.

\subsection{Photons as Euclidean co\-adjoint orbits}\label{Photons}

Let us start with a Lie group $G$ whose Lie algebra is denoted by $\fg$. Call $\vartheta$ the left-invariant Maurer-Cartan $1$-form of $G$. Let us then fix a point $\mu_0$ in the dual, $\fg^*$, of $\fg$, and define $\varpi=\mu_0\cdot\vartheta$, a preferred $1$-form of $G$. It is a classical result that the exterior derivative, $\sigma=d\varpi$, descends to the coadjoint orbit $\cO_{\mu_0}=G/\ker(\sigma)$ as the Kirillov-Kostant-Souriau symplectic $2$-form, $\omega$, which is, hence, canonically associated to $\mu_0\in\fg^*$.

\goodbreak

Here, we specialize this construction to the group of orientation-preserving iso\-metries of Euclidean space $E^3$, associated to $(\bbR^3, \la\,\cdot,\,\cdot\,\ra)$. Those constitute the Euclidean group, $\SE(3)$, iso\-morphic to the multiplicative group of the matrices
\begin{equation}
g=
\left(\begin{matrix}
R&\bx\\0&1
\end{matrix}
\right)
\label{Euclid}
\end{equation}
where $R=(\bu\,\bv\,\bw)\in\SO(3)$ is thought of as an orthonormal, positively oriented, frame of Euclidean space $E^3$, and $\bx\in(\bbR^3,+)$ is a translation.

\goodbreak

The $1$-form on $\SE(3)=\SO(3)\ltimes\bbR^3$ we consider, namely $\varpi=\mu_0\cdot{}g^{-1}dg$, is characterized by the generic Euclidean coadjoint-invariants $p>0$, and $s\neq0$ interpreted as the \textit{color} and the \textit{spin} \cite{Sou,GS,MR}. The explicit expression we will be using stems from the choice $\mu_0=(\bell_0,\bp_0)\in\se(3)^*\cong\bbR^3\times\bbR^3$, with 
\begin{equation}
\bell_0=
\left(
\begin{matrix}
s\\0\\0
\end{matrix}
\right)
\qquad
\&
\qquad
\bp_0=
\left(
\begin{matrix}
p\\0\\0
\end{matrix}
\right)
\label{mu0}
\end{equation}
so that $\varpi=\la\bell_0,j^{-1}(R^{-1}dR)\ra+\la\bp_0,R^{-1}d\bx\ra$, where $j:(\bbR^3,\times)\to\so(3)$ is the canonical Lie algebra isomorphism (with $\times$ the the standard cross-product). We, hence, find
\begin{equation}
\varpi=p\la\bu,d\bx\ra-s\la\bv,d\bw\ra.
\label{varpi0}
\end{equation}

\goodbreak

The $1$-forms (\ref{varpi0}) associated with \textit{photons}
correspond to the special values
\begin{equation}
s=\chi\hbar,
\label{spin}
\end{equation}
of spin, where $\hbar$ stands for the reduced Planck constant, and
\begin{equation}
\chi=\sign\,s
\label{chi}
\end{equation}
for the \textit{helicity}. See \cite{Sou,MR} and~\cite{DHH1,DHH2}. We will, in the sequel, denote these $1$-forms by
$\varpi_\chi=p\la\bu,d\bx\ra-\chi\hbar\la\bv,d\bw\ra$.
Notice that the choice (\ref{spin}) is dictated by the fact that geometric quantization of the model (see Section~\ref{GQ}) should ultimately lead to the Maxwell equations that rule wave optics. But the model just introduced should, nevertheless, be considered purely ``classical''.

\goodbreak

Let us now introduce the complex $3$-vector
\begin{equation}
\bz=\frac{\bv+i\bw}{\sqrt{2}}
\label{z}
\end{equation}
which clearly satisfies\footnote{The Euclidean scalar product, $\la\,\cdot\,,\,\cdot\,\ra$, of $\bbR^n$ is readily $\bbC$-linearly extended to 
$\bbC^n$, for all $n\in\bbN$.}
\begin{eqnarray}
\la\bz,\bz\ra&=&0,\label{z2=0}\\
\la\barbz,\bz\ra&=&1,\label{barzz=1}\\
\la\bu,\bz\ra&=&0,\label{uz=0}
\end{eqnarray}
where $\bz\mapsto\barbz$ stands for complex conjugation.
We recall that $\SO(3)$ is diffeomorphic to the submanifold $Z\hookrightarrow\bbC^3$ defined by Equations (\ref{z2=0}) and~(\ref{barzz=1}); this diffeomorphism 
\begin{equation}
{\mathsf{Z}\stackrel{\cong}{\longrightarrow}\SO(3)}:\bz\mapsto(\bu\,\bv\,\bw)
\label{Z}
\end{equation}
is given by Equation (\ref{z}) and
\begin{equation}
\bu=\frac{\barbz\times\bz}{i},
\label{u}
\end{equation}
where $\times$ is the standard cross-product of $\bbC^3$. The familiar principal fibration $\SO(3)\to{}S^2:\bz\mapsto\bu$ is therefore as in (\ref{u}).


We thus obtain the new expression of the $1$-form (\ref{varpi0}) on the Euclidean group, viz.,
\begin{equation}
\medbox{
\varpi_\chi=p\,\la\bu,d\bx\ra-\frac{\chi\hbar}{i}\la\barbz,d\bz\ra
}
\label{varpi}
\end{equation}

Let us recall that $d\varpi_\chi$ actually passes to Souriau's $5$-dimensional evolution space\footnote{The equivalence relation on $\SE(3)$ is given by $(\bz',\bx')\sim(\bz,\bx)$ iff $\bz'=e^{i\theta}\bz$ for some $\theta\in\bbR$, and $\bx'=\bx$.}
\begin{equation}
\cV=\SE(3)/\SO(2)\cong{}S^2\times\bbR^3
\label{V0}
\end{equation}
as a rank-$4$ presymplectic $1$-form $\sigma_\chi$; that is $d\varpi_\chi=(\SE(3)\to\cV)^*\sigma_\chi$. See Figure \ref{Diagram1}.

\goodbreak

Upon defining the new complex $3$-vector $\be_\chi=(\bv+i\chi\bw)/\sqrt{2}$, we find the useful decomposition 
\begin{equation}
\be_\chi=(\bz\ \barbz)\,\psi_\chi
\qquad
\hbox{where}
\qquad
\psi_\chi=
\left(
\begin{array}{c}
\half(1+\chi)\\[6pt]
\half(1-\chi)
\end{array}
\right)
\label{echi}
\end{equation}
on the basis $(\bz\ \barbz)$ of the complex $2$-space $(\bu^\perp)^\bbC$. These vectors, $\be_\chi$, also trivially satisfy Equations~(\ref{z2=0})--(\ref{uz=0}), where~$\chi=\pm1$ is the helicity (\ref{chi}). In view of~(\ref{z}), we have $\be_+=\bz$, and $\be_-=\barbz$. 
We can, hence, rewrite Equation (\ref{varpi}) as
\begin{equation}
\varpi_\chi=p\,\la\bu,d\bx\ra-\frac{\hbar}{i}\la\barbe_\chi,d\be_\chi\ra,
\label{varpichi}
\end{equation}
with
\begin{equation}
s\bu=\frac{\hbar}{i}\,\barbe_\chi\times\be_\chi,
\label{uchi}
\end{equation}
where $s=\chi\hbar$ denotes the photonic spin, and $\bu$ is as in (\ref{u}).

\goodbreak

The \textit{equations of motion} are associated with the characteristic foliation of the exact $2$-form~$d\varpi_\chi$ of $\SE(3)$, the $1$-form $\varpi_\chi$ being given by (\ref{varpi}); we easily get 
\begin{equation}
\delta(\bx,\bz)\in\ker(d\varpi_\chi)
\Longleftrightarrow
\left\{
\begin{array}{rcll}
\delta\bx&=&\alpha\,\bu\\
\delta\bz&=&i\beta\,\bz
\end{array}
\right.
\label{kerdvarpi}
\end{equation}
where $\alpha,\beta\in\bbR$ are Lagrange multipliers. 
As a direct consequence, and in view of~(\ref{u}), the \textit{direction of propagation}, $\bu$, is a constant of the motion (independent of $\chi$); the same is true for the photon \textit{location} $\bq=\bx-\bu\la\bu,\bx\ra\in\bu^\perp$. This entails that the coadjoint orbit $\cO_{\mu_0}=\SE(3)/\ker(d\varpi_\chi)\hookrightarrow\se(3)^*$ is diffeomorphic to the tangent bundle, $TS^2$, of the $2$-sphere~$S^2$ described by the pairs $(\bq,\bu)$. Hence
\begin{equation}
\cM=\cO_{\mu_0}\cong{}TS^2
\label{TS2}
\end{equation}
may be interpreted as the \textit{space of motions} (or of classical states) of photons of color $p$ and helicity $\chi$; this Euclidean coadjoint orbit is endowed with its canonical symplectic structure, $\omega_\chi$, viz., $d\varpi_\chi=(\SE(3)\to{}TS^2)^*\omega_\chi$, which reads
\begin{equation}
\omega_\chi=-p\,d\la\bq,d\bu\ra-\chi\hbar\,\bsurf_\bu,
\label{omega}
\end{equation}
where $\bsurf$ stands for the surface element of $S^2$, i.e.,
$
\bsurf_{\bu}(\delta\bu,\delta'\bu\ra=\la\bu,\delta\bu\times\delta'\bu\ra
$
for all vectors $\delta\bu,\delta'\bu\in{}T_\bu{}S^2$. See \cite{Sou}. Let us emphasizes that the Euclidean coadjoint orbit of photonics states $(\cM,\omega_\chi)$ of helicity $\chi$ possesses a \textit{Pukanszky polarization} \cite{DET,BGBR}, namely the vertical polarization; this entails automatically the ``twisted'' symplectic structure (\ref{omega}).

\subsection{Prequantization of colored photonic states}\label{Prequantization}

Putting $\bp=p\bu$, for the \textit{linear momentum}, elementary manipulations on Equation (\ref{varpichi}) then yield
$
\varpi_\chi=-\la\bx,d\bp\ra-\chi\hbar\,\la\barbzeta,d\bzeta\ra/i
$,
where
\begin{equation}
\displaystyle
\bzeta=e^{-\frac{i}{\chi\hbar}\la\bp,\bx\ra}\,\bz.
\label{zeta}
\end{equation}


Straightforward calculation shows that the $1$-form $\varpi_\chi$, given by (\ref{varpichi}), descends as
\begin{equation}
\alpha_\chi=-\la\bq,d\bp\ra-\frac{\chi\hbar}{i}\la\barbzeta,d\bzeta\ra
\label{alpha}
\end{equation}
on the circle-bundle $\cP=\SE(3)/(\ker({\varpi_\chi)}\cap\ker{(d\varpi_\chi}))$ over $\cO_{\mu_0}$. We also readily find that $d\alpha_\chi=(\cP\to\cO_{\mu_0})^*\omega_\chi$. 
See Figure \ref{Diagram1}.
One can show that
$\cP\cong{}TS^2\times_{S^2}\SO(3)$; at last $(\cP,\alpha_\chi)$ prequantizes $(\cM,\omega_\chi)$ in the sense of \cite{Kos,Sou}.

The last term in the RHS of (\ref{alpha}) may be interpreted as the \textit{Berry connection} of the principal circle-bundle $\SO(3)\to{}S^2$ over the $2$-sphere of unit momenta \cite{Ber1,SW}. This connection happens, hence, to be \textit{built in} within our approach! 

\begin{figure}[h]
\begin{displaymath}
\xymatrix{
\SE(3)
\ar[d]_{\SO(2)}\strut
\ar[r]^{\ \ \ker({\varpi_\chi})\,\cap\,\ker({d\varpi_\chi})} 
& 
\ar[d]^{\rU(1)} 
\red{\cP}\\
\blue{\cV\cong{}S^2{}\times\bbR^3}
\ar[r]^{\ \ \ker({\sigma_\chi})} 
&
{\cM\cong{}TS^2}
}
\end{displaymath}
\caption{Evolution \blue{$(\cV,\sigma)$} \& prequantum \red{$(\cP,\alpha_\chi)$} bundles over space of states $(\cM,\omega_\chi)$}
\label{Diagram1}
\end{figure}

The quantity~$\bzeta$ can be thought of as representing the \textit{polarization} vector of the \textit{prequantum state} $(\bq,\bzeta)\in\cP$ of our photon. It is, via (\ref{zeta}), an elementary solution of the Helmholtz equation
$
(\Delta+k^2)\bzeta=0,
$
where $\Delta=\sum_{j=1}^3{(\partial/\partial x^j)^2}$ is the Laplace operator and\begin{equation}
k=\frac{p}{\hbar}
\label{k}
\end{equation}
the ``wave-number'' associated with the color invariant, $p$.\footnote{The ``reduced wavelength''
\begin{equation}
\lambdabar=\frac{\hbar}{p}
\label{lambdabar}
\end{equation}
appears as a small parameter involved in semi-classical approximations \cite{BB04,BNKH} of Maxwell's equations.}

In view of (\ref{z2=0}), i.e., $\la\bzeta,\bzeta\ra=0$, we \textit{call} these (prequantum) states \textit{circularly polarized}. (This is just an interpretation since there is yet no electric and magnetic field at work; see below, however.) We claim that ours bundle $(\cP,\alpha_\chi)$ represents the optical pre\-quantum states of color~$p$ and handedness $\chi$, namely $(\bq,\bzeta)$ with $\chi=1$ is a right-handed pre\-quantum state, whereas~$(\bq,\barbzeta)$ is its left-handed companion.

\subsection{Polarization of light \& geometric quantization}\label{GQ}

Let us now quantize the model $(\cP,\alpha_\chi)$ by constructing the wave functions associated with a chosen $\SE(3)$-invariant polarization \cite{Kos}. In doing so, we will follow, almost \textit{verbatim}, the quantization of the massless \& spin-one coadjoint orbits of the restricted Poincar\'e group originally due to Souriau --- see Equations (19.135)--(19.161) in \cite{Sou}.

\goodbreak

Fix the helicity $\chi=+1$ to begin with. The manifold $\cP$ turns out to carry a natural $\rU(1)$-action, $(\bq,\bzeta)\mapsto(\bq,e^{-i\theta}\bzeta)$, for all $e^{i\theta}\in\rU(1)$. This action preserves the $1$-form~$\alpha_+$ given by  (\ref{alpha}); its infinitesimal generator $\delta(\bq,\bzeta)=(0,-i\bzeta)$ satisfies $\alpha_+(\delta(\bq,\bzeta))=\hbar$. This $1$-form is therefore (up to an overall factor $\hbar$) a $\mathrm{U}(1)$-connection whose curva\-ture, $d\alpha_+$, descends as our symplectic form, $\omega_+$, on $\cM=\cP/\mathrm{U}(1)\cong\cO_{\mu_0}$. Hence, $(\cP,\alpha_+)$ prequantizes $(\cM,\omega_+)$. 

The polarization \cite{Kos} we choose now is the vertical polarization $\bu=\const$ of the symplectic manifold $(\cM=TS^2,\omega_+)$.\footnote{The term ``polarization'' makes, here,  reference to a maximal isotropic distribution of a given symplectic manifold; it should not be confused with what is called polarization in wave optics!} This polarization is $\SE(3)$-invariant. Its horizontal lift to $(\cP,\alpha_+)$ is given by the trace on $\cP$ of the mixed vertical and anti-holomorphic polarization $\cF=\bigoplus_{j=1}^3\bbR\,\partial/\partial{}q^j\bigoplus_{j=1}^3\bbC\,\partial/\partial\zeta^j$ of $T^*\bbR^3\times\bbC^3$ (see~(\ref{alpha})).


The associated ``wave-function'' $\Psi:\cP\to\bbC$ are, by definition, $\mathrm{U}(1)$-equivariant and constant along $\cF$; they are thus antiholomorphic homogeneous functions of degree one in the variables $\bzeta\in\bbC^3$, hence of the general form
\begin{equation}
\Psi_+(\bq,\bzeta)=\la \barbzeta,\bA(\bu)\ra,
\label{Psi+}
\end{equation}
where $\bA:S^2\to\bbC^3$ is an otherwise arbitrary function.

\goodbreak

Now, the constraint (\ref{uz=0}) shows that $\pi\bz=\bz$ where $\pi$ is the ortho\-gonal projector on~$\bu^\perp$, so that $\Psi_+(\bq,\bzeta)=\la\barbzeta,\pi\bA(\bu)\ra$. This entails that the function~%
$
\bA
$
can be, with no loss of generality, restricted by the condition
\begin{equation}
\la\bu,\bA(\bu)\ra=0,
\label{pA=0}
\end{equation}
and becomes a section of the complexified tangent bundle of $S^2$.
In view of~(\ref{zeta}), our wave-functions (\ref{Psi+}) can be pulled-back to $\SE(3)$ as the functions
\begin{equation}
\widehat{\Psi}_+(\bx,\bz)=\la\barbz,e^{\frac{i}{\hbar}\la\bp,\bx\ra}\bA(\bu)\ra,
\label{psi+}
\end{equation}
which are completely determined by the functions $\bA$ already introduced; let us recall that $\bp=p\bu$ in (\ref{psi+}), where $\bu$ is given by (\ref{u}) in terms of $\bz$.

\goodbreak

Notice that geometric quantization of $(\cP,\alpha_-)$, for the opposite helicity, should lead, \textit{mutatis mutandis}, to wave-functions of the form
\begin{equation}
\Psi_-(\bq,\bzeta)=\la\bzeta,\bA(\bu)\ra,
\label{Psi-}
\end{equation}
with the same constraint (\ref{pA=0}) as before. 
We have, similarly,
\begin{equation}
\widehat{\Psi}_-(\bx,\bz)=\la\bz,e^{\frac{i}{\hbar}\la\bp,\bx\ra}\bA(\bu)\ra.
\label{psi-}
\end{equation}

It should be pointed out that the wave functions $\Psi_\pm$, associated with each helicity, help us recover the \textit{vector potential} through the following decomposition in terms of right and left-handed modes, viz.,
\begin{equation}
\medbox{
e^{\frac{i}{\hbar}\la\bp,\bx\ra}\bA(\bu)
=\bz\,\widehat\Psi_+(\bx,\bz)+\barbz\,\widehat\Psi_-(\bx,\bz)
}
\label{A}
\end{equation}

The last step consists in singling out from (\ref{A}) the (partial) Fourier transform
\begin{equation}
\tbA(\bx)=\int_{S^2}{\!\!e^{\frac{i}{\hbar}\la\bp,\bx\ra}\bA(\bu)\,\bsurf_\bu}.
\label{FA}
\end{equation}

This $\bC^3$-valued function of Euclidean space trivially satisfies the Helmholtz equation, and the Lorentz gauge condition (see (\ref{pA=0})), namely
\begin{equation}
(\Delta+k^2)\tbA=0
\qquad
\&
\qquad
\Div\tbA=0.
\label{PDE}
\end{equation}
Let us posit now
$\bE=ik\tbA$, and $\bB=\rot\,\tbA$. The system (\ref{PDE}) can be recast into the following system of PDE, viz.,
\begin{eqnarray}
\label{rotE}
\rot\,\bE-ik\bB&=&0,\\
\label{rotB}
\rot\,\bB+ik\bE&=&0,
\end{eqnarray}
known as the \textit{stationary Maxwell equations} for a given wave-number $k$, already introduced in (\ref{k}). These PDE represent the equations of a first-quantized theory of spinoptics in vacuum.

\goodbreak

The general solution (\ref{FA}) of (\ref{PDE}) is, hence, a wave-packet of mono\-chromatic plane waves, given by (\ref{A}), whose amplitudes $\bA(\bu)\neq0$ are orthogonal to~$\bu$ (see~(\ref{pA=0})); each elementary plane wave carries a (unitary) complex \textit{polarization vector}
\begin{equation}
\be= \frac{\bA(\bu)}{\Vert\bA(\bu)\Vert},
\label{e}
\end{equation}
where $\Vert\bA(\bu)\Vert^2=\la\overline{\bA(\bu)},\bA(\bu)\ra$. The mapping $\bu\mapsto\be$ can be thought of as a section of the  unitary complexified tangent bundle of the (real) $2$-sphere.

\goodbreak

Let us recall \cite{BW,GS} that
\textit{circular polarization} cor\-responds to the extra constraint
\begin{equation}
\la\be,\be\ra=0,
\label{circPol}
\end{equation}
(see Equation (\ref{z2=0})), while \textit{linear polarization} is described by the condition
\begin{equation}
\barbe\times\be=0.
\label{rectPol}
\end{equation}
\textit{Elliptic polarization} is the generic case. At last, the \textit{helicity} (or \textit{handedness}) of the polarization state $\be\,(\mathrm{mod}\,\rU(1))$ is
\begin{equation}
\chi=\sign\frac{\la\bu,\barbe\times\be\ra}{i},
\label{LR}
\end{equation}
and clearly coincides with the above definition of helicity for photonic states, i.e., $\chi=+1$ if~$\be=\bz\,(\mathrm{mod}\,\rU(1))$, whilst $\chi=-1$ if $\be=\barbz\,(\mathrm{mod}\,\rU(1))$.

This administers the proof that the polarization of light admits a clear-cut inter\-pretation in terms of geometric quantization of the classical ``circularly polarized'' states of geo\-metrical spinoptics. This remark constitutes the cornerstone of our theory of polarized geometrical spinoptics.

\section{Symplectic description of free polarized light}\label{PolarizedLightSection}

The main goal of this section is to extend the symplectic model for photons, reviewed in Section \ref{Photons}, to the case of light rays endowed with an internal structure, i.e., with an arbitrary \textit{polarization}. Our Ansatz is, hence, spelled out in the next section.

\subsection{A new evolution space}\label{NewEvolutionSpace}
  
As polarization of light obviously departs from the geometric objects of known element\-ary classical systems, and borrows some quantum features from Maxwell's equations (see Section \ref{GQ}), we will resort to a natural ``Hermitian extension'' of the previous formalism for classical spinning Euclidean particles. This will help us enlarge the group to start with, so as to gain the right number of extra degrees of freedom to account polarization of light.

We will, hence, introduce the \textit{polarization} via a unitary vector, $\be$, in the complexified ``wave plane'', $T^\bbC_\bu{}S^2$, considering  Equation (\ref{e}) as a hint.

\goodbreak

So, let us start with the overgroup $\SU(3)\ltimes\bbC^3$ of $\SE(3)=\SO(3)\ltimes\bbR^3$. Recall that any element of $\SU(3)$ retains the form $g=(\bu\,\,\be\,\,\barbu\times\barbe)$ where $(\bu\ \be)$ form an orthonormal system of vectors in $\bbC^3$, namely such that $\la\barbu,\bu\ra-1=\la\barbe,\be\ra-1=\la\barbu,\be\ra=0$. Here $\times$ is the cross-product $\bbC$-linearly extended to~$\bbC^3$.
Introduce the surjection
$
\mathrm{pr}_1\times\Id:\SU(3)\ltimes\bbC^3\to{}S^5\times\bbC^3:
((\bu\,\,\be),\bx)\mapsto(\bu,\bx)
$
as well as the natural embedding
$\Re:S^2\times\bbR^3\hookrightarrow{}S^5\times\bbC^3$,
where $S^2$ (resp.~$\bbR^3$) consists of those $\bu\in\bbC^3$ such that $\barbu=\bu$ (resp. of those $\bx\in\bbC^3$ such that $\barbx=\bx$).

Define the pulled-back bundle
$
\tcV\coloneqq{}\Re^*(\SU(3)\ltimes\bbC^3)
$
over $\cV=S^2\times\bbR^3$;
being given by
\begin{equation}
\tcV=\{(\bu,\be,\bx)\in\bbR^3\times\bbC^3\times\bbR^3\strut\big\vert\la\bu,\bu\ra=1, \la\barbe,\be\ra=1,\la\bu,\be\ra=0\},
\label{cV0}
\end{equation}
it will serve as our \textit{new evolution space}. We will denote by $\widetilde{\Re}:\tcV\hookrightarrow\SU(3)\ltimes\bbC^3$ the associated embedding. 
Let us show that $\tcV$ is nothing but than an associated bundle over the spinoptics evolution space $\cV$ (see Equation (\ref{V0}), and Figure \ref{Diagram}). This constitutes our prime justification for Definition (\ref{cV0}).

\begin{figure}[h]
\begin{displaymath}
\xymatrix{
\relax
\widetilde{\strut\SE(3)}\ar@{=}[d] 
&
\red{\tcV}\ar@{=}[d] &\red{\tcM}\ar@{=}[d]\\
\SU(3)\!\ltimes\!\bbC^3
\ar[d]_{\mathrm{pr}_1\times\,\Id}\strut
&
\ar@{_{(}->}[l]_{\widetilde{\Re}}\ 
 \ar[d]_{S^3}\strut
 \red{UT^\bbC{}S^2\!\times\!\bbR^3}\ 
  \ar[r]^{\ker(\red{\tsigma})} 
 & 
 \red{PT^\bbC{}S^2\!\times_{S^2}\!\!TS^2}\ar[d]^{\bbC{}P^1} \\
S^5\!\times\!\bbC^3
&
\ar@{_{(}->}[l]_{\Re}\ 
 {\color{blue}S^2\!\times\!\bbR^3}\ \ar@{=}[d] 
   \ar[r]^{\ker(\blue{\sigma})}\strut
 & {\color{blue}TS^2}\ar@{=}[d] \\ 
&{\color{blue}\cV}&{\color{blue}\cM\cong\cO_{\mu_0}}
}
\end{displaymath}
\caption{Evolution space \red{$(\tcV,\tsigma)$} \& space of polarized light states \red{$(\tcM,\tomega)$}}
\label{Diagram}
\end{figure}

The evolution space is, actually, $\tcV=\bbR^3\times{}UT^\bbC{}S^2$. The latter factor is related to the original principal $\SO(2)$-bundle $\SO(3)\to{}S^2$ as we shall explain. Now, in view of Equations (\ref{u2=1})--(\ref{eu=0}), the typical fibre of unitary bundle $UT^\bbC{}S^2\to{}S^2$ consists of \textit{unitary} complex $2$-vectors $\be\in(\bu^\perp)^\bbC$; it is thus diffeomorphic to $S^3\subset\bbC^2$. 
We can also think of this bundle in a way that makes closer contact with the Euclidean group and the quantization procedure outlined in Section~\ref{GQ}.
Indeed, start with the product $\SO(3)\times{}S^3$, described by the pairs~$(\bz,\psi)$, where $\bz\in\bbC^3$ is such that (\ref{z2=0}) and~(\ref{barzz=1}) hold, and
\begin{equation}
\psi=
\left(
\begin{matrix}
\psi_+\\ \psi_-
\end{matrix}
\right)
\in\bbC^2
\label{psi}
\end{equation}
satisfies $\la\barpsi,\psi\ra=\vert\psi_+\vert^2+\vert\psi_+\vert^2=1$, that is $\psi\in{}S^3$. 

\goodbreak

Consider the $\rU(1)$-action on this direct product, viz.,
$(\bz,\psi_+,\psi_-)\mapsto(e^{-i\theta}\bz,e^{i\theta}\psi_+,e^{-i\theta}\psi_-)$, for all~$e^{i\theta}\in\rU(1)$. This action being free, the orbit space is the $5$-dimensional manifold\footnote{The $S^3$-bundle $\SO(3)\times_{\SO(2)}S^3$ over $S^2$ is 
\textit{associated} to the principal $\SO(2)$-bundle~$\SO(3)\to{}S^2$.}
$\SO(3)\times_{\SO(2)}S^3\cong{}UT^\bbC{}S^2$,
the projection $\SO(3)\times{}S^3\to{}UT^\bbC{}S^2$ being given by $(\bz,\psi)\mapsto(\bu,\be)$ with
\begin{equation}
\bu=\frac{\barbz\times\bz}{i}
\qquad
\&
\qquad
\be=\left(\bz\ \barbz\right)\,\psi.
\label{ue}
\end{equation}

Equations~(\ref{ue}) are the straightforward generalization to elliptic polarization of Equations~(\ref{u}) and~(\ref{echi}) associated with the photonic case. We, furthermore, check that all constraints (\ref{u2=1})--(\ref{eu=0}) are duly satisfied by $\bu$, and $\be$, as given by~(\ref{ue}).

We have just proved that the \textit{new} evolution space $\tcV$ is an associated $S^3$-bundle over the original spinoptics evolution space $\cV=S^2\times\bbR^3$, namely
\begin{equation}
\medbox{
\displaystyle
\tcV\cong\big(\SO(3)\times_{\SO(2)}S^3\big)\times\bbR^3
}
\label{tVbis}
\end{equation}

\goodbreak

Let us emphasize, at this point, that the \textit{Jones vector}, $\psi$, introduced in (\ref{psi}), is conceptual\-ly related to the (normalized) two-component wave-function, $\Psi$, given  by~(\ref{Psi+}) and~(\ref{Psi-}). Equations (\ref{ue}) express the well-known relationship between the \textit{polariza\-tion vector}, $\be$, for a given direction, $\bu$,  of propagation, and the Jones vector, $\psi$, via the above decomposition of the polarization vector on the basis~$(\bz\ \barbz)$ of circular polarization state vectors~\cite{Bli}.




\subsection{A presymplectic model of polarized light in vacuum}\label{PreSympPolLight}

Let us show that our new evolution space, $\tcV$, can be endowed with a canonical presymplectic structure. To this end, we choose to consider the \textit{same} photonic origin (\ref{mu0}) as before, namely
$\mu_0=(\bs_0,\bp_0)\in\so(3)\times\bbR^3\subset\su(3)\times\bbC^3$,
where 
$$
\bs_0=s\left(\begin{matrix}0&&\\&i&\\&&-i\end{matrix}\right)
\qquad
\&
\qquad
\bp_0=\left(\begin{matrix}p\\0\\0\end{matrix}\right)
$$
with 
${s=\hbar}$ 
\& $p>0$. This natural choice prompts us to define a distinguished $1$-form 
on $\tcV$ as follows. If $\varTheta$ is the left-invariant Maurer-Cartan $1$-form of $\SU(3)\ltimes\bbC^3$, let us posit
\begin{equation}
\tvarpi=\widetilde{\Re}{\strut}^*(\mu_0\cdot\varTheta).
\label{tvarpi}
\end{equation}

The evolution space of polarized light rays, with color $p$, is therefore defined as the pair $(\tcV,\tsigma=d\tvarpi)$, where some calculation using (\ref{tvarpi}) shows that
\begin{equation}
\medbox{
\displaystyle
\tvarpi=p\la\bu,d\bx\ra-\frac{\hbar}{i}\la\barbe,d\be\ra
}
\label{tvarpiBis}
\end{equation}

The dynamics of polarized light rays takes place on the $8$-dimensional manifold, $\tcV$, described by the triples $y=(\bx,\bu,\be)$ where $\bx,\bu\in\bbR^3$, and $\be\in\bbC^3$ satisfy
\begin{eqnarray}
\la\bu,\bu\ra&=&1,\label{u2=1}\label{u2=1Bis}\\
\la\barbe,\be\ra&=&1,\label{baree=1}\\
\la\bu,\be\ra&=&0.\label{eu=0}
\end{eqnarray}


This new \textit{evolution space}, $\tcV$, according to the terminology of \cite{Sou}, is therefore endowed with the ``natural'' $1$-form (\ref{tvarpiBis})
that clearly restores the original $1$-forms $\varpi_\chi$, given by (\ref{varpichi}), once induced on the sub\-manifold $\SE(3)\hookrightarrow\tcV$ defined by $\be=\be_\chi$ (see (\ref{echi})).

\goodbreak

Call now $\hcV=\SE(3)\times{}S^3$ the \textit{extended evolution space}. Straightforward calcula\-tion shows that the pull-back $\widehat{\varpi}=(\hcV\to\tcV)^*\tvarpi$ of the $1$-form $\tvarpi$ on $\tcV$, introduced in~(\ref{tvarpiBis}), features the remarkable expression
\begin{equation}
\medbox{
\widehat{\varpi}=p\,\la\bu,d\bx\ra-\frac{\ts}{i}\la\barbz,d\bz\ra-\frac{\hbar}{i}\la\barpsi,d\psi\ra
}
\label{varpiTer}
\end{equation}
where 
\begin{equation}
\ts=\hbar\left(\vert\psi_+\vert^2-\vert\psi_-\vert^2\right)
\label{s3}
\end{equation}
is the ``third component'' of the \textit{Stokes vector},
$\bs=\hbar\,\bsigma^{-1}\left(2\psi\,\psi^*-\bone_{\bbC^2}\right)$,
that encodes the \textit{polarization state} of the system which is thereby  interpreted as a point on the \textit{Poincar\'e sphere} $S^2\cong\bbC{}P^1=\{\psi\,\psi^*\,\vert\,\psi\in{}S^3\subset\bbC^2\}$.\footnote{From now on, and wherever necessary, we use the notation $\psi^*=\la\barpsi,\,\cdot\,\ra$ for the adjoint of~$\psi$, say. As usual, we will also denote by $\bsigma=(\sigma_1,\sigma_2,\sigma_3)$ the Pauli matrices;
with this notation, we have $\ts=\hbar\,\psi^*\sigma_3\psi$.}

\goodbreak

We record, for future usage, that the new quantity $\ts$, defined by (\ref{s3}), and which shows up in~(\ref{varpiTer}), actually descends to the evolution space, $\tcV$, as the \textit{spin function}\footnote{Definition (\ref{s3}) readily authorizes the following interpretation (see Section \ref{GQ}), namely
\begin{itemize}
\item
$\ts=\pm\hbar$: right/left-handed \textit{circular} polarization,
\item
$\ts=0$: \textit{rectilinear} polarization,
\item
$\ts\in(-\hbar,0)\cup(0,+\hbar)$: generic \textit{elliptic} polarization.
\end{itemize}
}
\begin{equation}
\medbox{
\ts=\frac{\hbar}{i}\la\bu,\barbe\times\be\ra
}
\label{s3Bis}
\end{equation}
that enjoys the following property, viz.,
\begin{equation}
\ts\in[-\hbar,+\hbar].
\label{range(s)}
\end{equation} 
The novel \textit{spin} observable, $\ts$, given in (\ref{s3Bis}), whose sign, $\chi$, is the helicity (\ref{LR}) clearly replaces, in our approach the the spin $\SE(3)$-invariant, $s=\chi\hbar$, appearing in spinoptics (compare Equations (\ref{varpiTer}) and (\ref{varpi})). It naturally shows up, in this formalism, in the wake of the Ansatz (\ref{tvarpi}); see also (\ref{tvarpiBis}).

\goodbreak

Most interestingly, the equation (\ref{uchi}) relating the direction of propagation, $\bu$, to the circular polarization vector, $\be_\chi$, of light admits, in the present case, the quite similar form
\begin{equation}
\ts\,\bu=\frac{\hbar}{i}\,\barbe\times\be,
\label{spinVector}
\end{equation}
where the spin $\ts$ is, this time, given by (\ref{s3Bis}). 

It is worth noticing, at this point, that our $1$-form (\ref{tvarpiBis}) quite remarkably \textit{unifies}, in a single geometrical expression, namely (\ref{varpiTer}),
\begin{itemize}
\item
the \textit{Fermat contact $1$-form}: $\widehat{\varpi}_\mathrm{F}=\strut\la\bu,d\bx\ra$ on the photonic evolution space $\cV=\bbR^3\times{}S^2$,
\item
the \textit{Berry connection form}: $\widehat{\varpi}_\mathrm{B}=\displaystyle\frac{1}{i}\la\barbz,d\bz\ra$ on the circle-bundle $\SO(3)\to{}S^2$,
\item
the \textit{Pancharatnam connection form}: $\widehat{\varpi}_\mathrm{P}=\displaystyle\frac{1}{i}\la\barpsi,d\psi\ra$ on the Hopf bundle $S^3\to\bbC{}P^1$. 
\end{itemize}
We refer to \cite{Ber1,Pan,Ber2,SW} for the relevant original publications, and to \cite{Bli,BNKH} for a detailed review of the semi-classical Lagrangian description of wave optics, where the Berry correction term, and the polarization dynamics are introduced  in a somewhat independent fashion.

\goodbreak

\subsection{The states of polarized light as a symplectic manifold}\label{SpaceOfStates}

The equations of motion of polarized light are governed, as in (\ref{kerdvarpi}) by the characteristic foliation of $\tsigma=d\tvarpi$; the latter is given by a simple calculation, namely
\begin{equation}
\delta(\bx,\bu,\be)\in\ker(\tsigma)
\Longleftrightarrow
\left\{
\begin{array}{rcll}
\delta\bx&=&\alpha\,\bu,\\
\delta\bu&=&0,\\
\delta\be&=&i\beta\,\be,
\end{array}
\right.
\label{kerdvarpiBis}
\end{equation}
where $\alpha,\beta\in\bbR$ are Lagrange multipliers associated with the constraints~(\ref{u2=1Bis})--(\ref{eu=0}). 

\goodbreak

This entails that the $6$-dimensional symplectic manifold $\tcM=\tcV/\ker(\tsigma)$ is described by the triples $(\bq,\bu,\bpi)$ where $\bq=\bx-\bu\la\bu,\bx\ra\in\bu^\perp$ denotes, as for photons, the location of the light ray whose direction is $\bu\in{}S^2$, and $\bpi=\be\,\be^*\in{}PT_\bu^\bbC{}S^2$ stands for the Hermitian, rank-one, projector defining its polarization state.\footnote{We denote by $PT^\bbC{}S^2$ the projectivization of the complexified tangent bundle, $T^\bbC{}S^2=TS^2\otimes\bbC$, of the $2$-sphere, $S^2$; it is a $\bbC{}P^1$-bundle over $S^2$.}

Some more effort is needed to reveal the expression of the symplectic $2$-form, $\tomega$, of $\tcM$, such that $\tsigma=(\tcV\to\tcM)^*\tomega$; one finds
\begin{equation}
\medbox{
\tomega=-p\,d\la\bq,d\bu\ra-\frac{\hbar}{2i}\Tr\left(\bpi\left[d\bpi,d\bpi\right]\right)
}
\label{omegaBis}
\end{equation}
The symplectic manifold $(\tcM,\tomega)$ represents the \textit{space of motions} of (free) polarized light rays of fixed color $p$. One then readily shows that 
$
\tcM=\{(\bq,\bu,\bpi)\in\bbR^3\times\bbR^3\times{}L(\bbC^3)\,\strut\vert\,\break\la\bq,\bu\ra=0,\la\bu,\bu\ra=1,\bpi^2=\bpi=\bpi^*,\Tr(\bpi)=1,\bpi(\bu)=0\}
$
is indeed diffeomorphic to the $6$-dimensional fibered-product 
\begin{equation}
\medbox{
\tcM=TS^2\times_{S^2}PT^\bbC{}S^2
}
\label{tcM}
\end{equation}
above~$S^2$; the first factor represents the manifold of photonic states (see Section \ref{Photons}), whereas the novel second factor describes the polarization states of light in vacuum.

Let us emphasize, at this point, that the construct outlined in Section \ref{PreSympPolLight} does not hint at an obvious structure of $G$-homogenous symplectic manifold structure for~$(\tcM,\tomega)$, with~$G$ some natural overgroup of $\SE(3)$. The question as to whether the classical states of polarized light rays might be symplectomorphic to an elementary system of some finite-dimensional kin overgroup of the Euclidean group remains a compelling issue.

\subsection{The Euclidean momentum mapping}\label{J}

It is worth mentioning that the $1$-form $\tvarpi$, given by (\ref{tvarpiBis}) on the presymplectic manifold $\tcV$, is $\SE(3)$-invariant. 
This can be checked using the canonical lift to $\tcV$ of the $\SE(3)$-action on Euclidean space, namely $h_\tcV:(\bx,\bu,\be)\mapsto(A\bx+\bb,A\bu,A\be)$ where $h=(A,\bb)\in\SE(3)$; see~(\ref{Euclid}). One trivially has $h_\tcV^*\tvarpi=\tvarpi$, for all $h\in\SE(3)$.

A classical result states that this action canonically defines a ``momentum mapping'' $\sfJ:\tcV\to\se(3)^*$, given by the simple expression $\tvarpi(\delta_Z(y))=\sfJ(y)\cdot{}Z$, where~$\delta_Z$ stands for the fundamental vector field on $\tcV$ associated with $Z\in\se(3)$. This mapping, $\sfJ$, is moreover an integral invariant of the foliation $\ker(\tsigma)$ --- a fact which constitutes Souriau's pre\-symplectic formulation of Noether's theorem \cite{Sou}.

\goodbreak

Let us look for the explicit form of the momentum mapping $\sfJ=(\bell,\bp)$, using the natural pairing formula $\sfJ\cdot{}Z=\la\bell,\bomega\ra+\la\bp,\bbeta\ra$, for all $Z=(\bomega,\bbeta)\in\bbR^3\ltimes\bbR^3\cong\se(3)$.
We readily find $\delta_Z(\bx,\bu,\be)=(\bomega\times\bx+\bbeta,\bomega\times\bu,\bomega\times\be)$, so that we get $\varpi(\delta_Z(\bx,\bu,\be))=\sfJ(\bx,\bu,\be)\cdot{}Z$, where
\begin{eqnarray}
\label{bell}
\bell&=&\bx\times\bp+\ts\,\bu,\\
\label{bp}
\bp&=&p\,\bu,\label{p}
\end{eqnarray}
are the components of the $\SE(3)$-momentum mapping which represent, respectively, the \textit{angular momentum} and the \textit{linear momentum} of the model, the spin $\ts$ being as in (\ref{s3Bis}). These quantities descend as bona fide functions on the space of motions, $\tcM$.

\goodbreak

As a consequence, the observable $\ts$, appearing in (\ref{bell}), inherits (along with the color $p=\Vert\bp\Vert>0$)
the status of an $\SE(3)$-invariant of the extended model via its expression
$\ts=\Vert\bp\Vert^{-1}\la\bell,\bp\ra$;
it is also, again via the Noether theorem, a constant of the motion.

\subsection{The spin as a $\rU(1)$-momentum mapping}\label{U1Section}

There exists still another noteworthy group of automorphisms of $(\tcV,\tsigma)$, of the model of polarized light we have introduced. It is the group~$\rU(1)$ whose action on evolution space reads
$(e^{i\theta})_\tcV:(\bx,\bu,\be)\mapsto(\bx,\bu,\cos\theta\,\be+\sin\theta\,\bu\times\be)$. This group action on $\tcV$ stems from the following $\rU(1)$-action on $\hcV$, viz., $(\bx,\bz,\psi)\mapsto(\bx,e^{-i\theta}\bz,\psi)$, for all $e^{i\theta}\in\rU(1)$. 

\goodbreak

We easily check that this $\rU(1)$-action commutes with that of $\SE(3)$ on $\tcV$; it furthermore preserves the $1$-form $\tvarpi$ given by (\ref{tvarpiBis}), namely $(e^{i\theta})_\tcV^*\tvarpi\equiv\tvarpi$. The arguments of the preceding section apply just as well, insuring the existence of a $\rU(1)$-momentum mapping we are going to work out explicitly.
The infinitesimal action of this symmetry group is given by $\delta_\alpha(\bx,\bu,\be)=(0,0,\alpha\,\bu\times\be)$, where $\alpha\in\bbR$. 

Straightforward calculation then leaves us with $\tvarpi(\delta_\alpha(\bx,\bu,\be))=\sfS(\bx,\bu,\be)\cdot\alpha$, for all $\alpha\in\bbR$, where the $\rU(1)$-momentum mapping
\begin{eqnarray}
\label{J=s}
\sfS(\bx,\bu,\be)&=&\frac{\hbar}{i}\la\bu,\barbe\times\be\ra\\[6pt]
&=&\hbar\,\Tr(i\,\bpi\cdot{}j(\bu))
\label{J=sBis}
\end{eqnarray}
is nothing but the new spin observable, $\ts$, introduced in (\ref{s3Bis}); it conspicuously passes to the quotient $\tcM=\tcV/\ker(\tsigma)$ dealt with in Section \ref{SpaceOfStates}.

\goodbreak

We then notice that the space of polarized free light rays admits a double fibration,  given respectively (with a slight abuse of notation) by $\sfJ:\tcM\to\sfJ(\tcM)\subset\se(3)^*$ and $\sfS:\tcM\to[-\hbar,+\hbar]\subset\bbR$, viz.,
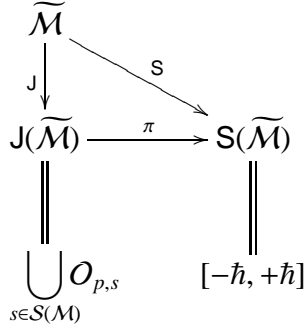
\begin{figure}[h]
\begin{displaymath}
\xymatrix{
& \tcM \ar[d]_\sfJ 
\ar@{>}[dr]^{\sfS} 
\\
& \sfJ(\tcM)\ar@{=}[d] \ar[r]^\pi & \ar@{=}[d]\sfS(\tcM)
\\
&\displaystyle\ \ \ \ \ \bigcup_{s\in\cS(\cM)}\!\!\!\cO_{p,s}&[-\hbar,+\hbar]
}
\end{displaymath}
\caption{Double fibration of space $\tcM$ of polarized light rays}
\label{DoubleFibration}
\end{figure}
above unions of coadjoint orbits of, respectively, $\SE(3)$ and $\rU(1)$.

The image of the Euclide\-an momentum mapping hence gives rise to a fibre bundle 
$\pi:\sfJ(\tcM)\to\sfS(\tcM):(\bell,\bp)\mapsto\ts$, whose fibres are $\SE(3)$-coadjoint orbits of color $p$ and spin $\ts\in[-\hbar,+\hbar]$ given by (\ref{J=s}) or (\ref{J=sBis}); see also (\ref{range(s)}), and Figure \ref{DoubleFibration}.

\goodbreak

\section{Polarized spinoptics in inhomogeneous media}\label{InHomogeneousSection}

Geometrical optics in inhomogeneous/anisotropic dielectric media has long proved crucial in theoretical and experimental optics; its basic principles mainly rely on Riemann/Finsler, as well as symplectic geo\-metry~\cite{Arn,Sou,AIM}. 
For example, the \textit{Fermat principle} amounts to considering light rays as unparametrized geodesics of a Riemannian manifold $(M,\rg)$, where~$M$ is a smooth open submanifold of $\bbR^3$, and
\begin{equation}
\rg=n(\bx)^2\delta_{ij}\,dx^i\otimes{}dx^j
\label{FermatMetric}
\end{equation}
the so-called ``Fermat metric'' as\-sociated with the ``refractive index'' $n\in{}C^\infty(M,\bbR^*_+)$ of the in\-homogeneous and isotropic optical medium under consideration. (Einstein's summation convention is used in Equation (\ref{FermatMetric}), and throughout this article.)

This viewpoint has then been extended to \textit{geometrical spinoptics} in an effort to deal with spinning photons in arbitrary isotropic \cite{DHH1,DHH2}, and anisotropic \cite{Duv} media.
It has first been suggested, in these references, to resort to the procedure of \textit{minimal coupling} (borrowed from general relativity) to a Riemannian metric, in order to axiomatize the (pre)symplectic formulation of the dyna\-mics of light rays in a background Riemannian metric, e.g., the ``Fermat metric''~(\ref{FermatMetric}).

\subsection{Polarized spinoptics in a Riemannian manifold}\label{RiemannCase}

Let us, hence, start with an arbitrary connected, orientable, smooth $3$-dimensional Rieman\-nian manifold $(M,\rg)$, whose metric is denoted by
\begin{equation}
\rg=\rg_{ij}(X)\,dX^i\otimes{}dX^j.
\label{g}
\end{equation}
in a local coordinate system $(X^1,X^2,X^3)$. Let us call $\nabla$ the Levi-Civita connection of $(M,\rg)$, and $\Vol=\sqrt{\det(\rg_{ij}(X))}\,dX^1\wedge{}dX^2\wedge{}dX^3$ its canonical Riemannian volume form. 

Then, the $8$-dimensional evolution space we will introduce, in the very spirit of Section~\ref{PolarizedLightSection}, is defined by
\begin{equation}
\tcV=\{(X,U,E)\in{}TM\times_M{}T^\bbC\!M\,\vert\,\rg(U,U)=1,\rg(U,E)=0,\rg(\barE,E)=1\}
\label{cV}
\end{equation}
in terms of constraints that faithfully reproduce those given in (\ref{u2=1Bis})--(\ref{eu=0}).

\goodbreak

We must invoke, at this stage, the procedure of minimal coupling which dictates that the Euclidean metric, $\la\,\cdot,\,,\,\cdot\,\ra$, be replaced by the Riemannian metric, $\rg$, and the differential, $d$, be supplanted by the covariant differential, $d^\nabla$, of tensor fields when passing from flat Euclidean space to a curved (pseudo-)Riemannian manifold.\footnote{See, e.g., \cite{Duv} for another approach using the bundle of oriented, orthonormal frames of $(M,\rg)$.} 

The $1$-form of $\tcV$ to consider in place of (\ref{tvarpiBis}), for polarized light of color $p$, reads then\footnote{We systematically employ a useful and shorthand notation such as $\rg(U,dX)=\rg_{ij}(X)U^idX^j$.}
\begin{equation}
\medbox{
\tvarpi=p\,\rg(U,dX)-\frac{\hbar}{i}\rg(\barE,d^\nabla\!E)
}
\label{varpiNew}
\end{equation}
with the (local) expression of the covariant differential, viz., $d^\nabla\!E^\ell=dE^\ell+\Gamma_{jk}^\ell{}E^jdX^k$, for all $\ell=1,2,3$; the $\Gamma_{jk}^\ell$ denote, here, the Christoffel symbols of $\nabla$ in the chosen coordinate patch.


Just as before, we claim that the dynamics of the system is governed by the characteris\-tic foliation of the $2$-form $\tsigma=d\tvarpi$ of $\tcV$. Call $y=(X,U,E)$, so that easy calculation yields
\begin{eqnarray}
\tsigma(\delta{y},\delta'{y})&=&p\left[\rg(\delta^\nabla\!U,\delta'X)-\rg(\delta'^\nabla\!U,\delta{X})\right]
-\frac{\hbar}{i}\rg(\barE,R(\delta{X},\delta'{X})E)\nonumber\\[6pt]
&&-\frac{\hbar}{i}\left[\rg(\delta^\nabla\barE,\delta'^\nabla\!E)-\rg(\delta'^\nabla\barE,\delta^\nabla\!E)\right]
\label{sigmaNew}
\end{eqnarray}
for all $\delta{y},\delta'{y}\in{}T_y\tcV$. We note in (\ref{sigmaNew}) the appearance of the Riemann curvature tensor, $R$, via its defining relationship $R(\delta{X},\delta'{X})E\equiv\delta^\nabla\delta'^\nabla\!E-\delta'^\nabla\delta^\nabla\!E-[\delta,\delta']^\nabla\!E$.

\goodbreak

Let us now determine the characteristic foliation of the $2$-form $\tsigma$ as given by~(\ref{sigmaNew}) on~$\tcV$ (see (\ref{cV})). 
We thus have to reveal the form of all $\delta{y}\in{}T_y\tcV$ such that $\tsigma(\delta{y},\delta'{y})+\delta'\left[\alpha\rg(U,U)+\beta\rg(\barE,E)+\bargamma\rg(U,E)+\gamma\rg(U,\barE)\right]=0$ for all~$\delta'{y}\in{}T_y\tcV$, where $\alpha,\beta\in\bbR$, and $\gamma\in\bbC$. 
One gets immediately
\begin{eqnarray}
p\,\delta{X}&=&\alpha{}U+\bargamma{}E+\gamma\barE\label{deltaX},\\[6pt]
p\,\delta^\nabla\!U&=&\frac{\hbar}{i}R(E,\barE)\delta{X},\label{deltaU}\\[6pt]
\frac{\hbar}{i}\delta^\nabla\!E&=&-\beta{}E-\gamma{}U.\label{deltaE}
\end{eqnarray}
We must then impose that these equations be actually compatible with the above constraints. One has, indeed, $\delta(\rg(U,U))=0$ whenever $\mathrm{Im}(\bargamma{}R(E,\barE,U,\barE))=0$, i.e., $\gamma=\varrho{}R(E,\barE,U,\barE)$ for some $\varrho\in\bbR$.\footnote{We have put $R(E,\barE,U,\barE)\equiv\rg(\barE,R(E,\barE)U)$.} The second constraint is trivially satisfied, $\delta(\rg(\barE,E))=0$. As for the third one, $\delta(\rg(U,E))=0$, one readily finds $\gamma/\hbar+(\hbar/p)R(E,\barE,\delta{X},E)=0$; this, together with Equation (\ref{deltaX}), entails that the new parameter is  given by $\varrho=\alpha(\hbar/p)^2\left[1-(\hbar/p)^2R(E,\barE,E,\barE)\right]^{-1}$, and that~$\gamma$ is therefore expressed in terms of the (real) Lagrange multiplier $\alpha$. 

We can now claim that
\begin{equation}
\begin{array}{c}
\delta(X,U,E)\in\ker(\tsigma)\\[10pt]
\Updownarrow\\[12pt]
\left\{\begin{array}{lcl}
\delta{X}
&=&
\alpha\left[
\displaystyle
U
+
\frac{\lambdabar^2}{\Delta}\Big(E\,R(E,\barE,U,\barE)-\barE\,R(E,\barE,U,E)\Big)\right],
\\[12pt]
\delta^\nabla\!U
&=&
\displaystyle
\frac{\lambdabar}{i}R(E,\barE)\delta{X},
\\[10pt]
\delta^\nabla\!E
&=&
i\beta{}E
-
\displaystyle
\alpha{}\frac{\lambdabar}{i\Delta}\,U\,R(E,\barE,U,E),
\end{array}
\right.
\end{array}
\label{kersigmaNew}
\end{equation}
for some (redefined) $\alpha,\beta\in\bbR$, and where
$\Delta=1-\lambdabar^2R(E,\barE,E,\barE)$
is generically nonzero. 

\goodbreak

Notice, in (\ref{kersigmaNew}), the polarization-induced anomalous velocity of order $\cO(\lambdabar^2)$ where $\lambdabar$ stands for the reduced wavelength (\ref{lambdabar}). Of course, the 
general 
folia\-tion (\ref{kersigmaNew}) helps us recover, in the \textit{flat} case, the characteristic foliation (\ref{kerdvarpiBis}) of the ``free'' system.

\goodbreak

The definition (\ref{s3Bis}) prompts the following intrinsic definition\footnote{Note that $\sfS:\tcV\to\bbR$ defined by $\sfS(X,U,E)=\ts$, as given by Equation (\ref{sNew}), is nothing but the $\rU(1)$-momentum mapping of Section \ref{U1Section}, which still exists in the general Riemannian case; its range is, again, $\sfS(\tcV)=[-\hbar,+\hbar]$.}
\begin{equation}
\ts=\frac{\hbar}{i}\,\Vol(U,\barE,E)
\label{sNew}
\end{equation}
of the spin, which readily turns out to be a conserved quantity, i.e., an integral invariant of the foliation (\ref{kersigmaNew}), namely
\begin{equation}
\delta{\ts}=0.
\label{deltas=0}
\end{equation}
The (formal) quotient $\tcM=\tcV/\ker(\tsigma)$ thus inherits a structure of $6$-dimensional symplectic manifold, interpreted as the \textit{space of motions}, or the space of polarized optical states, of color~$p$, in the Riemannian manifold $(M,\rg)$.

\goodbreak

\subsection{Polarized spinoptics in a Fermat manifold}

Let us now specialize, in the above calculation, the Riemannian metric (\ref{g}) to the 
Fermat metric (\ref{FermatMetric}) on $M\subset{}E^3$.

\goodbreak

The definition (\ref{cV}) of evolution space, $\tcV$, prompts us to define the quantities~$\bx$, $\bu$, and $\be$ via $x^j=X^j$, $u^j=nU^j$, and $e^j=nE^j$ for all $j=1,2,3$; those turn out to satisfy (\ref{u2=1})--(\ref{eu=0}), as required. From the familiar expression of the Christoffel symbols, namely $\Gamma_{ij}^k=n^{-1}(\delta^k_i\partial_jn+\delta^k_j\partial_in-\delta^{k\ell}\delta_{ij}\partial_\ell{}n)$, for all~$i,j,k=1,2,3$, one straightforwardly  gets the covariant differential $d^\nabla\!E^k=n^{-1}(de^k+n^{-1}(e^j\partial_jn)dx^k-n^{-1}(\delta^{k\ell}\partial_\ell{}n)\delta_{ij}e^idx^j)$, for all $k=1,2,3$.

With these preparations, our $1$-form (\ref{varpiNew}) can easily be rewritten in the following guise, viz.,
$
\varpi
=
n\la{}p\bu,d\bx\ra
-\hbar/(in)\left[
n^{-1}\la\barbe,d\be\ra+n^{-2}\la\barbe,d\bx\ra\,\be(n)-n^{-2}\la\be,d\bx\ra\,\barbe(n)\ra
\right]
$,
with the notation $\be(n)=e^j\partial_jn$. 

\goodbreak

Defining, for convenience,
$
\bg=\grad(n^{-1}),
$
some more calculation leads us to the $1$-form 
\begin{equation}
\varpi=\la\bp,d\bx\ra-\frac{\hbar}{i}\la\barbe,d\be\ra
\label{Fermat1form}
\end{equation}
where
\begin{equation}
\bp=n \Big(p \bu+\frac{\hbar}{i}\bg\times(\barbe\times\be)\Big)
\label{Fermatp}
\end{equation}
is the brand-new \textit{canonical} ``momentum'' of the theory.

The evolution space $(\tcV,\tvarpi)$ is best defined in terms of the triples $(\bx,\bp,\be)\in{}M\times\bbR^3\times\bbC^3$ subject to the constraints
\begin{eqnarray}
\Vert\bp\Vert^2
&=&
n^2(p^2+\hbar^2\Vert\bg\times(\barbe\times\be)\Vert^2),\label{p2New}\\
\la\barbe,\be\ra
&=&1,\label{barbebeNew=1}\\
\la\bp,\be\ra
&=&
in\hbar\la\bg,\be\times(\barbe\times\be)\ra\label{peNew}.
\end{eqnarray}

The explicit determination of the characteristic foliation of the $2$-form $\tsigma=d\tvarpi$ of $\tcV$, using the above constraints, would need a somewhat involved calculation yielding, ultimate\-ly, the \textit{exact} equations of polarized light rays in an isotropic, inhomogeneous, dielectric medium. In order to allow comparison with other approaches~\cite{BB04,OMN}, we will, instead, confine considerations to the case of a \textit{slowly} varying refractive index, $n$, dealt with in those references. We will thus determine the sought characteris\-tic foliation up to terms of order $\cO(\Vert\bg\Vert^2)$ and~$\cO(\Vert\partial\bg/\partial\bx\Vert)$. Hence, suffice it to replace Equation~(\ref{p2New}) by 
\begin{equation}
\Vert\bp\Vert^2\approx{}n^2p^2,
\label{p2NewApprox}
\end{equation}
keeping the other constraints (\ref{barbebeNew=1}) and (\ref{peNew}) unchanged.

\goodbreak

Much in the same way than in Section \ref{RiemannCase}, we find that $\delta(\bx,\bp,\be)\in\ker(\sigma)$ if 
\begin{eqnarray}
\delta\bx&\approx&\alpha\bp+\bargamma\be+\gamma\barbe\label{deltax},\\
\delta\bp&\approx&-\alpha{}p^2n^3\bg,\label{deltap}\\
\delta\be&\approx&i\beta\be-\frac{i}{\hbar}\gamma\bp,\label{deltae}
\end{eqnarray}
where $\alpha,\beta\in\bbR$, and $\gamma\in\bbC$ are Lagrange multipliers associated to the constraints~(\ref{p2NewApprox}), (\ref{barbebeNew=1}), and (\ref{peNew}) respectively.

\goodbreak 

Compatibility of the distribution (\ref{deltax})--(\ref{deltae}) with these constraints henceforth yields $\gamma=i\alpha{}n\hbar\la\be,\bg\ra$, implying $\gamma=\cO(\Vert\bg\Vert)$. At last, we get
\begin{equation}
\delta(\bx,\bp,\be)\in\ker(\tsigma)
\Longleftrightarrow
\left\{
\begin{array}{lcl}
\delta\bx
&\approx&
\alpha\Big[
\displaystyle
\bp
+
\frac{\hbar}{i}n(\barbe\times\be)\times\bg\Big],
\\[10pt]
\delta\bp&\approx&-\alpha{}p^2n^3\bg,
\\[10pt]
\delta\be&\approx&i\beta\be+\alpha{}n\bp\la\be,\bg\ra,
\end{array}
\right.
\label{kersigmaNewFermat}
\end{equation}
where $\alpha,\beta\in\bbR$.

\goodbreak

The equation for the velocity in (\ref{kersigmaNewFermat}) can be easily recast in the following form
\begin{equation}
\medbox{
\delta\bx
\approx
\alpha\Big[
\displaystyle
\bp
+
\frac{\ts}{p}\bp\times\bg\Big]
}
\label{deltasxTer}
\end{equation}
which, again, highlights the presence of an anomalous velocity, transverse to the gradient of the (slowly) varying refractive index, and proportional to the spin (\ref{s3Bis}) of the polarization. Let us stress that Equation (\ref{deltasxTer}) exactly matches Equation (5) in \cite{BNKH}; see also~\cite{BB04,OMN}. We refer to \cite{DHH1,DHH2} for the case of photonic spinoptics.

Easy calculation shows, moreover, that the spin (\ref{s3Bis}) is, indeed, an integral invariant of the foliation (\ref{kersigmaNewFermat}), i.e.,
\begin{equation}
\medbox{
\delta{\ts}\approx0
}
\label{deltas=0Bis}
\end{equation}
in full agreement with the with the general result (\ref{deltas=0}), and notably with those of~\cite{Bli,BNKH} (preces\-sion of the Stokes vector) obtained using a different approach based on a semi-classical approximation of stationary wave optics.


\subsection{The Spin Hall Effect of Light}\label{SHEL}

Let us finish with the geometric derivation of the spin-induced modification of the Snel-Descartes laws \cite{Des,BW} of reflection/refraction of light rays through a sharp  dielectric interface; this goes under the names of ``Optical Hall Effect'', or ``Spin Hall Effect of Light'' (SHEL) in the optics literature. We contend that the SHEL, although pertaining to Maxwell's wave optics theory, may nevertheless be understood within the theory of polarized spinoptics.

Consider a plane in Euclidean space, $E^3$, separating two half-spaces $M_1$ and~$M_2$ which are endowed with a Fermat metric (\ref{FermatMetric}) associated with constant refractive indices~$n_1$ and~$n_2$ respectively.\footnote{Our approach may  account for negative indices characterizing the newly discovered meta\-materials \cite{Pen}; we refer to \cite{DHH1} for a plausible explanation of the \textit{perfectness} of metamaterial lenses in terms of spinoptics.} Let us call $\bn$ the unit normal to the interface pointing towards $M_2$, say.
Since the ordinary differential equations (\ref{kersigmaNewFermat}) governing light ``propagation'' in a smooth refractive index are clearly inapplicable in this situation, we need to resort to a scattering theory adapted to such a classical set-up. To this end, we choose Souriau's  principle of ``symplectic scattering'' \cite{Sou}, for which we now provide a brief account.

Start with two symplectic manifolds $(\cM_1,\omega_1)$ of ``in'' and $(\cM_2,\omega_2)$ of ``out'' classical free states of a given system. Consider that the correspondence between these spaces of asymptotic states is governed by a scattering process which needs not be explicitly described. Souriau's fundamental assumption lies in the fact that the scattering is given by a local diffeo\-morphism
$\cS:\cM_1\to\cM_2$ such that $\omega_1=\cS^*\omega_2$. We will from now on consider the case where the asymptotic free states are (open subsets of) some Hamiltonian $G$-spaces. To take account of the symmetries of the scattering device through the subgroup $H\subset{}G$ they are assumed to define, we will consistently look at those local scattering diffeomorphisms, $S$, intertwining the $H$-action, namely such that $\cS\circ{}h_{\cM_1}=h_{M_2}\circ\cS$ for all $h\in{}H$. Call $\fg$ (resp.~$\fh$) the Lie algebra of $G$ (resp.~$H$). If $\sfJ_a:\cM_a\to\fg$ denote, for $a=1,2$, the corresponding momentum mapping, it is a trivial matter to check that there holds 
\begin{equation}
\sfJ_1|\fh=S^*(\sfJ_2|\fh)
\label{J1h=J2h}
\end{equation}
provided $\cM_1$ and $\cM_2$ are connected \cite{Sou,DHH1}. The ``conservation law'' (\ref{J1h=J2h}) will play a central r\^ole in the sequel.

Let us apply this to the case where $\cM_a$ is the open submanifold of $\tcM$ (see (\ref{tcM})) consis\-ting of those rays, in $M_a$, crossing the optical interface, viz., of the $y_a=(\bq_a,\bu_a,\bpi_a)\in\tcM$ such that $\la\bu_a,\bn\ra\neq0$, for each $a=1,2$. In view of Equations (\ref{Fermat1form}) and (\ref{Fermatp}), the color in the half-space $M_a$ is $p_a=p\,n_a$, for $a=1,2$, where $p$ is the original color Casimir in vacuum.

Now, the symmetry group of the considered plane interface is clearly $H=\SE(2)$, i.e., the subgroup of $G=\SE(3)$ generated by rotations around the normal $\bn$, and translations in~$\bn^\perp$. Taking advantage of the results of Section \ref{J}, we claim that the $H$-momentum mappings read $\sfJ_a=(L_a,\bP_a)$ where $L_a=\la\bn,\bell_a\ra$, and $\bP_a=\bn\times\bp_a$ for $a=1,2$, together with Equations~(\ref{bell}) and~(\ref{bp}). Positing $(\bq_2,\bu_2,\bpi_2)=\cS(\bq_1,\bu_1,\bpi_1)$, we can read off the conserva\-tion law (\ref{J1h=J2h}) as follows, viz.,
\begin{eqnarray}
\label{L1=L2}
\la\bn,\bq_1\times\bp_1+\ts_1\bu_1\ra&=&\la\bn,\bq_2\times\bp_2+\ts_2\bu_2\ra\\
\label{P1=P2}
\bn\times\bp_1&=&\bn\times\bp_2
\end{eqnarray}
where $\bp_a=p_a\bu_a$, and 
$\ts_a=\hbar\,\Tr(i\,\bpi_a\cdot{}j(\bu_a))$, see (\ref{J=sBis}), for $a=1,2$. 

Equation (\ref{P1=P2}), i.e., $\bp_2=\bp_1+\lambda\bn$, where $\lambda$ is an explicit function of $\bp_1$, precisely corresponds to the familiar \textit{Snel-Descartes laws} \cite{BW}, either for reflection or for refraction.

As to Equation (\ref{L1=L2}), it easily yields $\la\bn,(\bq_2-\bq_1)\times\bp_1\ra=\ts_1\la\bn,\bu_1\ra-\ts_2\la\bn,\bu_2\ra$; now, positing quite generically $\bq_2=\bq_1+\mu\bp_1+\nu\bn+\varrho\bn\times\bp_1$, for some functions $\mu,\nu,\varrho$, we end up with $\la\bn,(\bq_2-\bq_1)\times\bp_1\ra=-\varrho\Vert\bn\times\bp_1\Vert^2=\ts_1\la\bn,\bu_1\ra-\ts_2\la\bn,\bu_2\ra$. 

\goodbreak

Let us single out the shift $\Delta\bq=\varrho\bn\times\bp_1$ between the location, $\bq_2$ of the outgoing light ray relatively to that, $\bq_1$, of the incoming one; this shift is thus transversal to the plane of incidence spanned by $\bp_1$ and $\bn$. In view of the preceding calculation we immediately obtain the expression of the \textit{transverse shift}, namely\footnote{It can be checked directly that the transverse shift actually vanishes at normal incidence.}
\begin{equation}
\bigbox{
\Delta\bq=\frac{\left[\ts_2\la\bn,\bu_2\ra-\ts_1\la\bn,\bu_1\ra\right]}{\Vert\bn\times\bp_1\Vert}\,\frac{\bn\times\bp_1}{\Vert\bn\times\bp_1\Vert}
}
\label{SHEL}
\end{equation}
in accordance with the formula of the SHEL originally found in \cite{BB04} and \cite{OMN}. There remains, however, to achieve the daunting computation of the sought scattering symplectomorphisms, $\cS$, to obtain explicitly the spin observable $\ts_2$ in terms of the incoming data $y_1=(\bq_1,\bu_1,\bpi_1)$.

Let us finally recall that the uniqueness of the above-mentioned symplectomorphism inter\-twining the $\SE(2)$-action has been established \cite{DHH1} for reflection/refraction in photonic spinoptics. We defer to subsequent work the proof of the (still conjectured) uniqueness of the reflection/refraction symplectomorphism in the more elaborate framework of polarized spinoptics.

%
%
%
%
%
%
 



\end{document}